# Charging performance of the Su-Schrieffer-Heeger quantum battery

Fang Zhao,[1,2] Fu-Quan Dou,[3,*] and Qing Zhao[1,4,†]

[1]*Center for Quantum Technology Research and Key Laboratory of Advanced Optoelectronic Quantum Architecture and Measurements (MOE), School of Physics, Beijing Institute of Technology, Beijing 100081, China*
[2]*China Academy of Engineering Physics, Beijing 100088, China*
[3]*College of Physics and Electronic Engineering, Northwest Normal University, Lanzhou 730070, China*
[4]*Beijing Academy of Quantum Information Sciences, Beijing 100193, China*

The Su-Schrieffer-Heeger (SSH) model has recently received considerable attention in condensed matter. Here, we investigate SSH-based charging protocols of quantum batteries (QBs) with $N$ spins. The hopping interaction induces ground state splitting, leading to different effects on the dimerization parameter to the QB in different quantum phase regimes. In the degenerate ground state regime, the dimerization parameter has little influence on the QB. In contrast, the dimerization parameter has a significantly quantum advantage in the fully nondegenerate ground state regime on energy and ergotropy. It also results in the dimerization spin pairs having larger occupations than other spins. Although the dimerization parameter can enhance the energy and ergotropy, the QB's capacity will decrease.

## I. INTRODUCTION

The quantum battery (QB) has emerged as a new research field and recently has attracted remarkable attention [1]. A QB is a quantum system that can potentially provide temporary energy storage. It was first proposed in a single two-level spin system that stores energy from an external field [2]. Up to now, the main interest in QBs has focused on the charging process, including how to increase the charging and releasing energy [3–16]. Since a QB's energy is usually released in a thermal heat bath, the QB's energy for work has been commonly discussed in terms of ergotropy [17–26]. Other researchers are also focused on the entanglement or work-extraction capability of QBs [22,27–35].

In the usual case, a QB is charged by an external field. An energy-charged cavity field in an excited energy state can save as much as an external field [3]. Energy oscillations during the charging process necessitate accurate control of the charging time. A magnetic field directly charges other charging sources [7,11,36] or is charged by a thermal heat bath [19]. To charge by a thermal bath, the discussion has to be centered on a QB in an open quantum system [19–26,37].

It has been verified that an $N$-spin chain coupled to a cavity field can significantly enhance the charging power of QBs [3,38]. However, the hopping interaction between each spin in a QB has seldom been taken into consideration [15,37–42]. The simplest interaction in a spin chain is the nearest hopping interaction. In our previous paper, we dealt with a spin chain with identical nearest hopping interactions between neighboring spins, which could significantly enhance the energy and ergotropy of QBs [42]. Here, we move on to a bit more complex type of interaction. A spin chain with a staggered hopping interaction, modeled by the Su-Schrieer-Heeger (SSH) model, will be considered [43–45]. The SSH model has been widely explored in the polyacetylene molecule. As the simplest one-dimensional (1D) model presenting topological behavior, recently it has been realized in many fields, such as circuit QED, cold atom, graphene nanoribbons, plasmonics, and photonics systems [46–52].

In this paper, we study the charging properties of QBs with the SSH spin chain model, as shown in Fig. 1. We focus on investigating the maximum charging energy and ergotropy of QBs. Here, the extra coherent driven field is the source to charge the SSH QB, which forces us to control the charging time of the QB to ensure maximum charging energy. We study the energy and ergotropy of QBs with the nearest-neighbor interaction and the dimerization parameter. The nearest-neighbor interaction of a QB could bring about a quantum phase transition (QPT), as discussed in our previous paper [42]. Thus we investigate SSH QBs in different quantum phase regions. Then we analyze how the dimerization parameter influences the energy and ergotropy of QBs in different regions. We also discuss the dimerization spin pairs and other spins in different regions. Finally, we study the dimerization parameter that influences the QB's capacity.

The rest of the paper is organized as follows. In Sec. II, we introduce the model of the SSH QB and define the energy and ergotropy of a QB. Then we discuss the charging process of a QB with and without dimerization parameters in Secs. III and IV, respectively. The conclusion is given in Sec. V.

*Corresponding author: doufq@nwnu.edu.cn
†Corresponding author: qzhaoyuping@bit.edu.cn

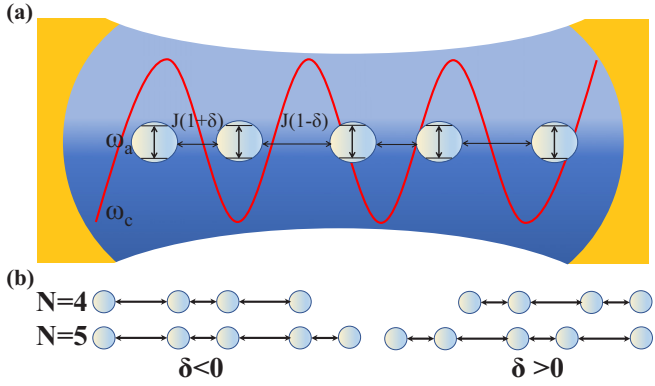

FIG. 1. (a) A schematic diagram of the SSH QB charging protocol. It includes an $N$ spin with a frequency of $\omega_a$. The spin chain has a nearest-neighbor hopping interaction with a strength of $J$. The $\delta$ is the dimerization parameter. The spin chain is coupled with a photon cavity with a frequency of $\omega_c$. (b) A schematic diagram of the SSH model spin chain with dimerization parameters $\delta < 0$ and $\delta > 0$ for spin numbers $N = 5$ and $N = 6$, respectively.

## II. MODEL

In this paper, we study the energy charging and release of a SSH QB. Figure 1(a) shows the charging protocol. The physical model of a SSH QB is an $N$-spin chain with frequency $\omega_a$. There is a hopping interaction between nearest-neighbor spins with strength $J$ and the dimerization constant is $\delta$. The QB is coupled with a cavity field with a cavity frequency of $\omega_c$. This cavity field transfers energy to charge the QB. Figure 1(b) shows the spin chain distribution for odd- and even-spin chains with different dimerization situations.

The whole Hamiltonian of this SSH QB charging system is

$$H_S = H_A + H_B + H_I.$$

$$H_A = \omega_c c^\dagger c,$$

$$H_B = \omega_a \sum_{i=1}^{N} \sigma_+^{(i)} \sigma_-^{(i)}$$

$$- J(1+\delta) \sum_{i=1,3,5,...}^{N-1} (\sigma_+^{(i)} \sigma_-^{(i+1)} + \text{H.c.})$$

$$- J(1-\delta) \sum_{i=2,4,6,...}^{N-2} (\sigma_+^{(i)} \sigma_-^{(i+1)} + \text{H.c.}),$$

$$H_I = \sum_{i=1}^{N} g(\sigma_+^{(i)} c + \text{H.c.}). \quad (1)$$

Here, $H_S$ is the total Hamiltonian of the whole QB charging system. $H_A$ is the cavity Hamiltonian. $\omega_c$ is the cavity field frequency. $H_B$ is the SSH QB Hamiltonian. $J$ is the nearest-neighbor hopping strength between spins. $\delta$ is the dimerization parameter. $H_I$ is the interaction term between the cavity and the QB. $g$ is the coupling constant between the spin and the cavity.

We consider the charging process of a SSH QB in a closed quantum system. Here, the initial state of the QB is the empty state

$$|V_0\rangle = |g\rangle_B, \quad (2)$$

where $|g\rangle$ is the ground state of the QB. The charging of the QB is to let the empty QB gain energy from the external cavity. In our charging protocol, the cavity is full of energy. When the cavity and QB interaction strength $g$ are nonzero, the QB will start charging. The dynamic charging process of the QB will be

$$|\psi(t)\rangle = U|\psi(0)\rangle = e^{-iH_S t}|\psi(0)\rangle. \quad (3)$$

$|\psi(0)\rangle$ is the initial state of the whole system,

$$|\psi(0)\rangle = |V_0\rangle \otimes |\phi\rangle. \quad (4)$$

$|\phi\rangle$ is the cavity state with full energy. We take the cavity initially in the Fock state,

$$|\phi\rangle = |n_c\rangle. \quad (5)$$

$n_c$ is the cavity photon number.

Energy storage during the QB at time $t$ is given by

$$E_B(t) = \text{tr}[H_B \rho_B(t)],$$

where $\rho_B(t) = \text{tr}_A[\rho_S(t)]$ is the reduced density matrix of the QB at time $t$. The energy charged into the QB is equal to $E_B(t) - E_B(0)$, where $E_B(0) = E_G$ is the ground state energy of the QB. Therefore, the actual charging energy of the QB is equal to

$$\Delta E(t) = E_B(t) - E_G. \quad (6)$$

The $\Delta E(t)$ is used to determine the energy charged into the QB. However, QB's energy $\Delta E(t)$ cannot be entirely released. According to the second law of thermodynamics, $\Delta E(t)$ cannot be wholly transformed into valuable work without dissipating heat. Therefore, the concept of ergotropy is introduced to characterize the ability to generate valuable work of a QB. The ergotropy is defined as

$$\varepsilon_B(t) = E_B(t) - \min_U \text{tr}[H_B U \rho_B(t) U^\dagger]. \quad (7)$$

The $H_B$ and $\rho_B$ can be diagonalized and represented as

$$\rho_B(t) = \sum_n r_n(t)|r_n(t)\rangle\langle r_n(t)|,$$

$$H_B = \sum_n e_n|e_n\rangle\langle e_n|.$$

The eigenvalues of $\rho_B(t)$ are arranged in descending order as $r_0 \geqslant r_1 \geqslant \cdots$, and the eigenvalues of $H_B$ are arranged in ascending order as $e_0 \leqslant e_1 \leqslant \cdots$. The second term on the right-hand side of the definition of ergotropy can be simplified as

$$\min_U \text{tr}[H U \rho(t) U^\dagger] = \sum_n r_n e_n. \quad (8)$$

## III. CHARGING PROPERTIES WITHOUT DIMERIZATION PARAMETER

In this section, we investigate the charging process of a QB without considering the dimerization parameter. Then we discuss how to determine the charging time to ensure the QB obtains the maximum energy and ergotropy. Except for the dimerization parameter $\delta$, the hopping strength $J$ still brings on the quantum phase (QPT) in the QB. Hence we also discuss the hopping parameter, and the ground state quantum phase influences the energy and the QB ergotropy. We discuss the odd and even chains, respectively, due to the SSH model spin chain's symmetry.

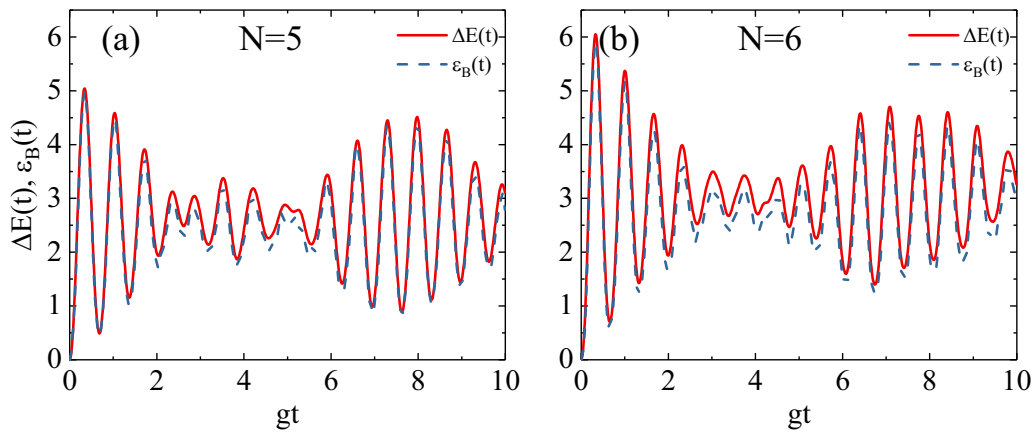

FIG. 2. The dynamic energy and ergotropy of a QB with different hopping strengths. (a) and (b) are the energies and ergotropies of a QB with spin numbers $N = 5$ and $N = 6$, respectively. Other parameters are $\omega_a = \omega_c = g = J = 1$, $\delta = 0$.

In our QB charging model, $\omega_a$ and $\omega_c$ must be equal to ensure maximum energy transfer. The parameter $g$ mainly influences the energy oscillator frequency. In all calculations, we take $\omega_a$ as a dimensionless parameter and let $\omega_a = 1$. For simplicity, other parameters are taken as $\omega_c = g = \omega_a = 1$. Thus, we only focus on the critical parameters $J$ and $\delta$. Owing to the different structures from the SSH spin chain with odd and even spins, we consider $N = 5$ and $N = 6$, respectively. To ensure the QB acquires enough energy and ergotropy, the source cavity photon number should be $N_c > 2N$. The real photon in the cavity source is the multiphoton state. The minimum photon number $N_c$ is required to simulate the charging throughout the cavity source to obtain accurate numerical calculation results. We assume the source cavity is in the Fock states. The numerical results of the charging process of a QB with time are shown in Fig. 2.

The energy and ergotropy oscillate between the spin and the outside cavity source. The dynamics of SSH QBs for the odd and even chains have similar properties. The maximum energy and ergotropy of a QB appear at the first peak during the charging process. For the maximum energy charging of a QB, we have to take this time as the charging time $\tau_c$. Thus the charging time to be controlled is $\tau_c \propto 1/\sqrt{N}$. In the following discussions, we consider only the QB's maximum energies and ergotropies that appear at the charging time $t = \tau_c$.

Now we focus on the hopping interaction in a QB. Here, we only discuss the maximum energy and ergotropy of a QB during the charging process. For the nondimerization situation ($\delta = 0$), the calculated maximum energies and ergotropies of a SSH QB with different hopping interactions are shown in Figs. 3(a) and 3(b).

Without considering the dimerization parameter, both the odd and even chains have similar behavior with the hopping interaction. The energy and ergotropy have some discontinuous points for particular hopping strengths. For further discussion, we calculate the energy spectrum, as shown in Figs. 3(c) and 3(d). Each discontinuous point of the energy and ergotropy corresponds to a ground energy state crossing point. These discontinuous points on the energy spectrum indicate a QPT. The SSH spin chain model may also lead to a topological phase transition [53]. However, in our QB charging model, the system's energy is mainly decided by the bulk state of the whole spin chain. Hence we only discuss the classical QPT.

To study the quantum phase here, we introduce two ordering parameters. One of the common ordering parameters is the $z$ component of the averaged magnetization $M_z$,

$$M_z = \frac{\langle S_z \rangle_g}{N}, \quad (9)$$

where $\langle \cdots \rangle_g$ represents taking the average on the ground state, and the total spin operator is $S_z = \sum_{i=1}^{N} \sigma_z^i$. Here, we define another ordering parameter $\xi_z$ as

$$\xi_z = \frac{\langle S_z^2 \rangle_g}{N^2}, \quad (10)$$

These parameters characterize the magnetic fluctuations in the spins along the $z$ axes.

The ordering parameters for odd and even chains are shown in Figs. 3(e) and 3(f). Before the first ground state energy crossing point, where the QB gets the maximum energy, the QB's ground state is in the ferromagnetic phase. After the first ground state energy crossing point, the energy band becomes split. Until the last ground state energy crossing point, the ground state energy band is fully split. In the next section, we focus on discussing the QB's maximum energy and ergotropy in the degenerate and fully nondegenerate ground state regimes.

## IV. CHARGING PROPERTIES WITH DIMERIZATION PARAMETER

### A. Dimerization effects

Owing to the difference between odd- and even-spin chains when considering the dimerization parameters, we here discuss them separately. For the odd-spin chain, the maximum

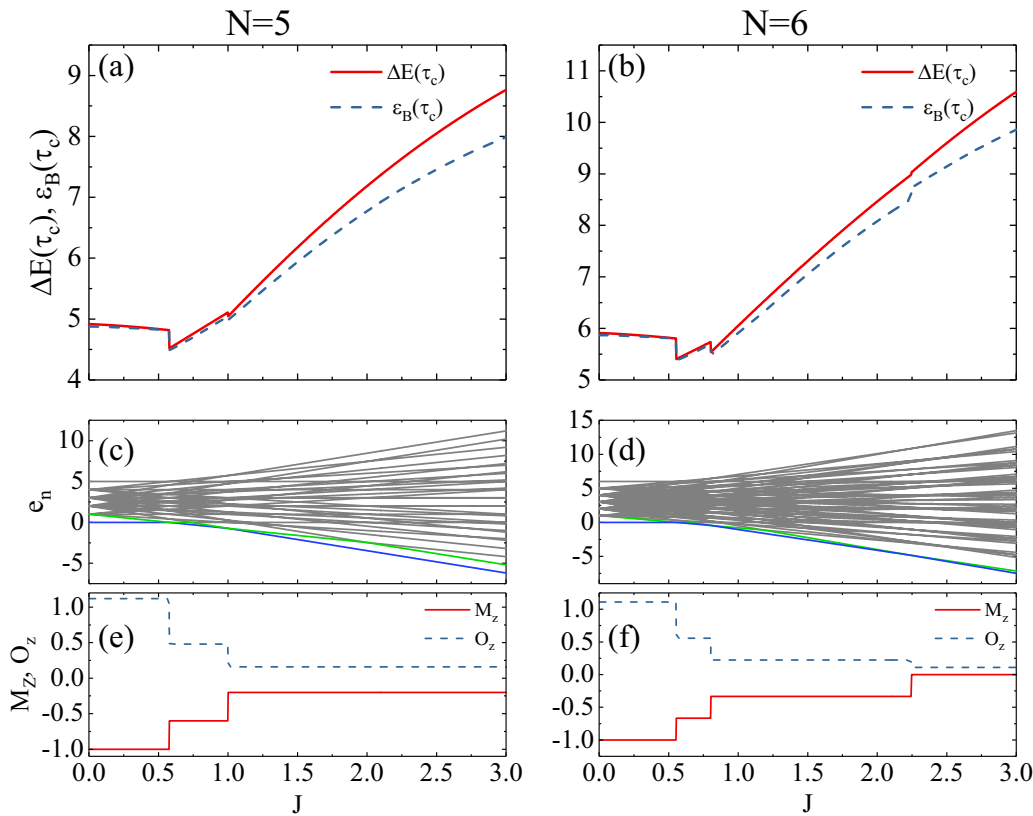

FIG. 3. The maximum energy and ergotropy, spectrum, and the ordering parameters of a QB with different hopping strengths. (a) and (b) are the energies and ergotropies of a QB with spin numbers $N = 5$ and $N = 6$, respectively. (c) and (d) are the spin numbers' respective energy spectra. (e) and (f) are the ordering parameters of a QB.

energies and ergotropies of a QB with different dimerization parameters are shown in Fig. 4.

In the degenerate ground state regime, the energy and ergotropy are hardly influenced by the dimerization parameter. A larger dimerization parameter leads to a lower energy and ergotropy for the QB. Owing to the odd number of spins, the plus and minus dimerization parameter will appear in the same number of spin pairs, as shown in Fig. 1(b). That makes the energy and ergotropy symmetrical with the dimerization parameters.

In the fully nondegenerate ground state regime, the dimerization parameter significantly influences the energy and ergotropy. A larger dimerization parameter leads to a much higher energy and ergotropy for the QB. The energy and ergotropy are also symmetrical with the dimerization parameter.

Now we consider the even-spin chain. The numerical results are shown in Fig. 5. In the degenerate ground state regime, the energy and ergotropy change in a very small range similar to the case of an odd-spin chain. However, there is also a clear difference to be noticed. The energy and ergotropy are no longer symmetrical with the dimerization parameters. There is one more pair of spins having an interaction of strength $J(1 + \delta)$ than that of $J(1 - \delta)$ in an even-spin chain. The ground state energy is lower when $\delta < 0$ and thus the charged energy $\Delta E$ is higher.

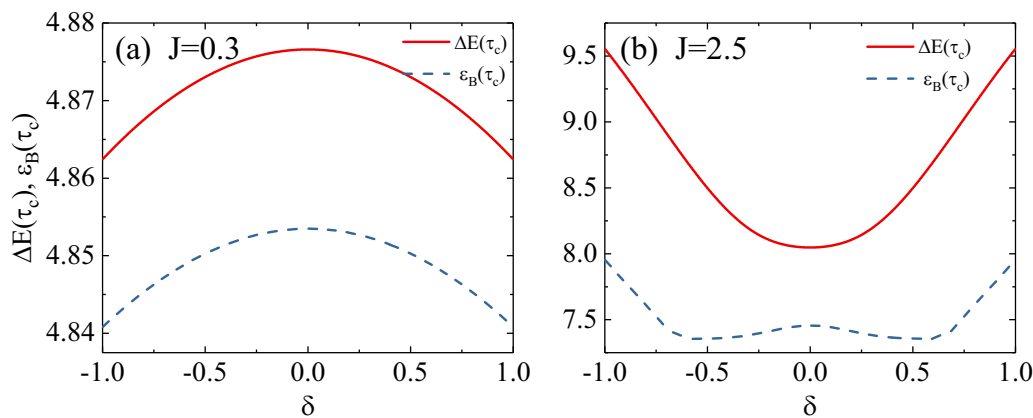

FIG. 4. The maximum energy and ergotropy of a QB with the spin number $N = 5$. (a) and (b) are the energies and ergotropies for $J = 0.3$ and $J = 2.5$, respectively.

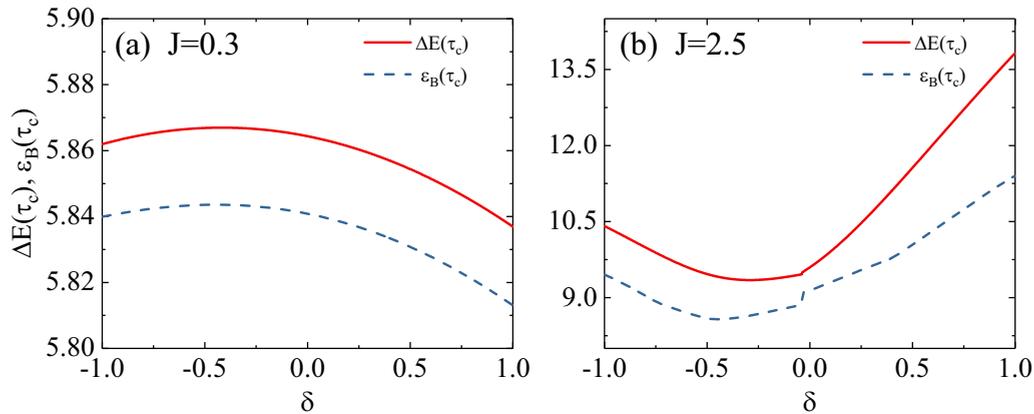

FIG. 5. The maximum energy and ergotropy of a QB with spin number $N = 6$. (a) and (b) are the energies and ergotropies for $J = 0.3$ and $J = 2.5$, respectively.

As for the fully nondegenerate ground state regime case with $N = 6$, the dimerization parameter significantly influences the energy and ergotropy. The larger dimerization parameter will bring a much higher energy and ergotropy than the nondimerization situation. The $\delta > 0$ that corresponds to a much higher energy and ergotropy is also due to the positive $\delta$ having one more pair of spins, as shown in Fig. 1(b).

We find that in both the odd- and even-spin chains, the dimerization parameter has the same effect. The energy and ergotropy have significantly increased in the fully nondegenerate ground state regime and slightly decreased in the degenerate ground state regime. For the extreme dimerization situation ($\delta = \pm 1$), the energy and ergotropy have been increased by about 15%–40%. In the degenerate ground state regime, the energy and ergotropy have been only decreased by about 0.3%.

We have already discussed the dimerization parameter in the degenerate and fully nondegenerate ground state regimes. Now we will calculate the maximum energy and ergotropy versus $J$ and $\delta$, as shown in Fig. 6. When the QB is in the degenerate ground state regime, the maximum energy and ergotropy hardly change with the dimerization parameter and the hopping interaction strength. However, in the fully nondegenerate ground state regime, the dimerization parameter will significantly improve the SSH QB's maximum energy and ergotropy. In this region, when there are more spin pairs with hopping interaction strengths larger than $J$ ($\delta > 0$), the energy and ergotropy become higher, as seen in the upper right-hand corners in Figs. 6(c) and 6(d).

In the other regions (the undiscussed quantum phase area), the SSH QB's ground state will frequently cross with other excited states. After each crossing point, the dimerization may influence the energy and ergotropy of the QB. The numbers of ground state crossing points increase with spin numbers, which makes it difficult to discuss the dimerization parameter in these regions in detail. Fortunately, the energy and ergotropy are not significantly influenced by the dimerization parameter in these regions.

### B. Quantum advantage of dimerization spin pairs

In the previous section, we have already discussed how the dimerization parameter influences the SSH QB. We have confirmed that the dimerization parameter will improve the QB's energy and ergotropy in the energy's fully nondegenerate area. Now we further investigate how the dimerization parameter improves the maximum energy and ergotropy of a QB.

Now we calculate the occupation of each spin. The occupation of the $i$th spin is defined as $O_i = \langle \sigma_+^i \sigma_-^i \rangle_{\max}$. $\langle \cdots \rangle_{\max}$ is averaged on the charged QB state. First, we calculate the occupation of each spin for the odd-spin chain ($N = 5$), as shown in Fig. 7. The occupation also slightly influences the degenerate ground state regime. Each dimerization spins in pairs, which corresponds to a nearly full occupation. Due to the degenerate ground state regime, the dimerization parameter will suppress the energy and ergotropy, making the dimerization spin pairs correspond to lower occupations. For the fully nondegenerate ground state regime, the dimerization parameter has a significant influence on the occupation. Each dimerization spins in pairs, which corresponds to a higher occupation. The minus and plus dimerization parameters appear in the numbers of the same spin pairs, which makes the occupation symmetrical with the dimerization parameter.

Then we calculate the occupation of the even-spin chain ($N = 6$), as shown in Fig. 8. The occupation also has a slight influence on the degenerate ground state regime. Each dimerization spins in pairs, which corresponds to a lower occupation. For the fully nondegenerate ground state regime, the occupation has a significant influence, as shown in Fig. 8(b). It is also different from the degenerate ground state regime. Each dimerization spins in pairs, which corresponds to a higher occupation.

From comparing the odd- and even-chain occupation, we have found that the occupation has the same influence as the dimerization parameters. For the degenerate ground state regime, the dimerization spin couples have a smaller occupation. The smaller energy and ergotropy mean that the dimerization suppresses the QB energy and ergotropy in the degenerate ground state regime. However, the dimerization spin pairs have a huge occupation with a larger energy and ergotropy in the fully nondegenerate ground state regime. That means the dimerization parameters are beneficial to the energy and ergotropy of a QB on the fully nondegenerate ground state regime.

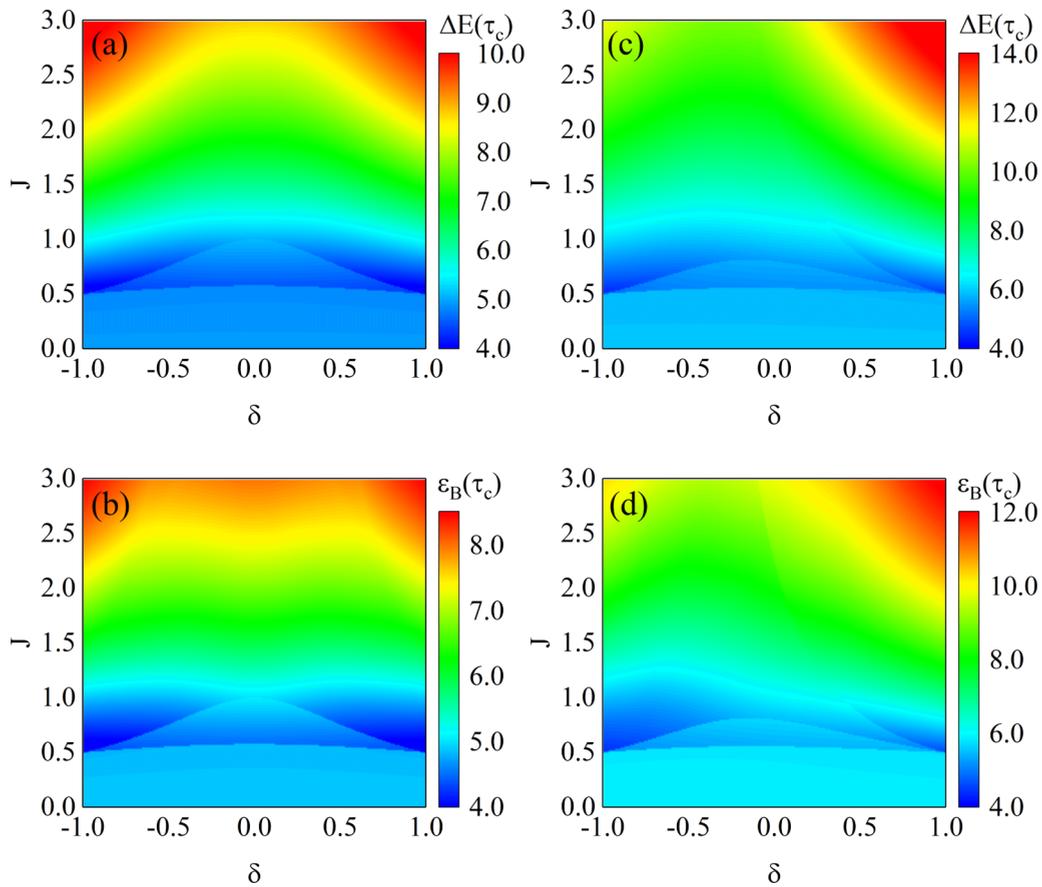

FIG. 6. The maximum energy and ergotropy of a QB. (a) and (c) are the QB's energy with the spin numbers $N = 5$ and $N = 6$, respectively. (b) and (d) are the QB's ergotropy with the spin numbers $N = 5$ and $N = 6$, respectively.

Now we define the parameter $R$ to characterize the charge capacity for the QB's energy and ergotropy. It is defined as follows:

$$R_{eb} = \frac{\Delta E}{E_{\max} - E_G}, \quad (11)$$

$$R_{epb} = \frac{\varepsilon_B}{E_{\max} - E_G}. \quad (12)$$

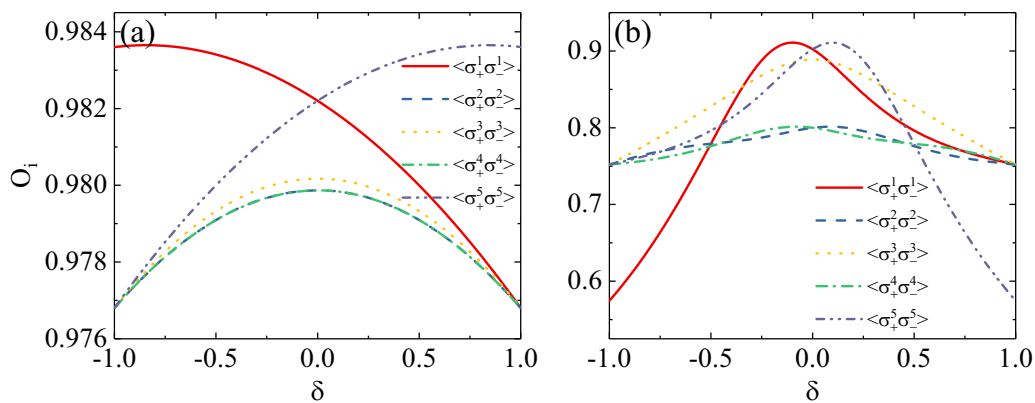

FIG. 7. The occupations of each spin with spin numbers $N = 5$. (a) and (b) are the spin occupations for hopping strengths $J = 0.3$ and $J = 2.5$, respectively.

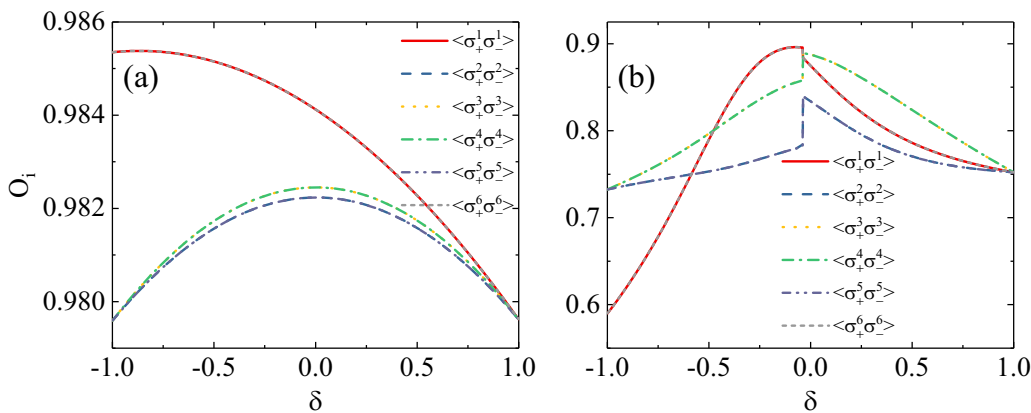

FIG. 8. The occupations for each spin with the spin numbers $N = 6$. (a) and (b) are the spin occupations for the hopping strengths $J = 0.3$ and $J = 2.5$, respectively.

$E_{max}$ is the highest-energy level of a QB and $E_G$ is the ground state energy. We now discuss the capacity of a QB with the dimerization parameter.

The capacities in the degenerate and the fully nondegenerate ground state regimes are shown in Fig. 9. We find that both in the degenerate and fully nondegenerate ground state regimes, the capacities $R_{eb}$ and $R_{epb}$ are decreased with the dimerization parameter $\delta$. In addition, the capacity is more sensitive in the fully nondegenerate than in the degenerate ground regime.

Owing to the numerical calculation restriction, we could only calculate the maximum spin number as $N = 6$. However, from the capacity of the calculated result of the QB, we find that in the fully nondegenerate ground state regime for the most dimerized institution $\delta = \pm 1$, the increase of the spin numbers will increase the QB capacity. It means

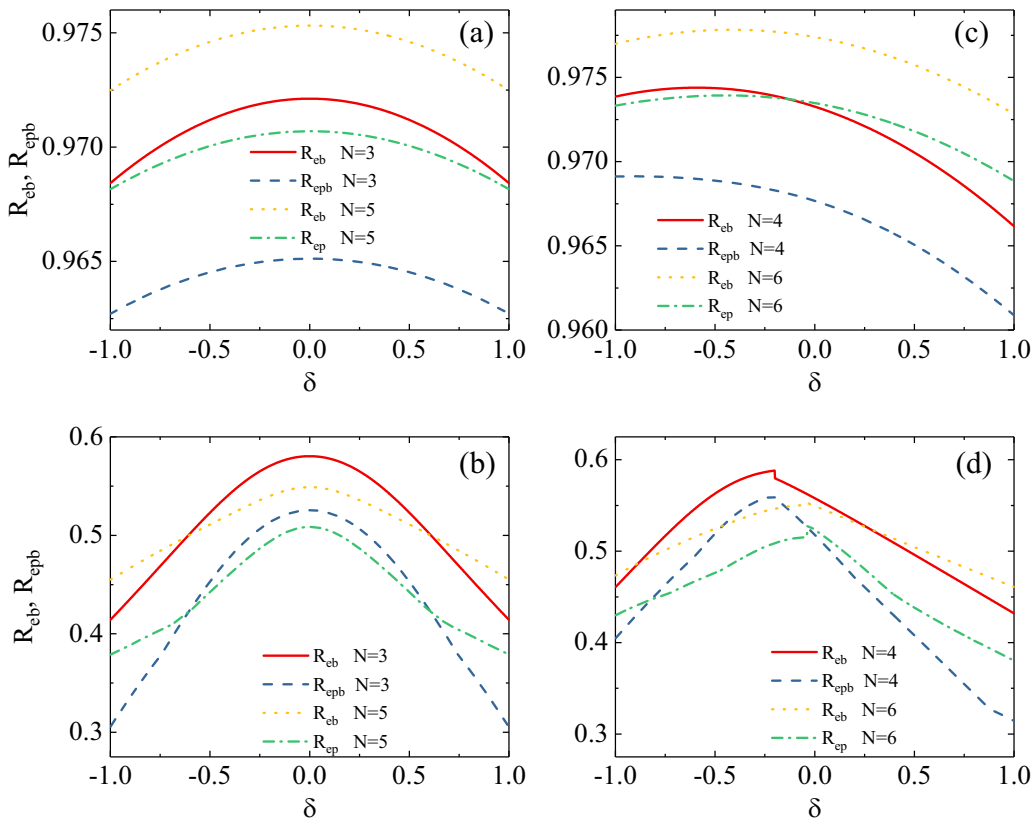

FIG. 9. The QB's capacity for odd and even chains. (a) and (c) are the capacities of odd and even chains in the degenerate ground state regime. (b) and (d) are the capacities of odd and even chains in the fully nondegenerate ground state regime. The other parameters are (a), (c) $J = 0.3$, and (b), (d) $J = 2.5$.

that we could enhance the capacity by increasing the spin numbers.

## V. CONCLUSION

In conclusion, we have investigated the SSH QB charging properties with an external cavity field. First, the maximum energy and ergotropy appear at the first peak during the charging process in our charging protocol, which leads to an optimal charging time $\tau_c \propto 1/\sqrt{N}$. In the degenerate ground state regime, the dimerization parameter has little influence on the energy and ergotropy. The large dimerization parame-ter corresponds to a relatively smaller energy and ergotropy. However, in the fully nondegenerate ground state regime, the large dimerization parameter will dramatically increase the QB's energy and ergotropy. This is because dimerization spin pairs have relatively larger occupations than other spins. Finally, although the dimerization parameter will enhance the energy and ergotropy, the QB's capacity decreases.

## ACKNOWLEDGMENTS

This work is supported by the National Science Foundation (NSF) of China with Grants No. 11675014 and No. 12075193.